\documentstyle[aps,psfig,multicol]{revtex}
\newcommand{\ibar}{\bar{\imath}}
\newcommand{\jbar}{\bar{\jmath}}
\newcommand{\kbar}{\bar{k}}
\begin{document}
\draft

\title{Dimensional crossover in half-filled and lightly doped $N$-leg Hubbard ladders}
\author{Urs Ledermann}
\address{Theoretische Physik, Eidgen\"{o}ssische Technische Hochschule,
CH-8093 Z\"{u}rich, Switzerland}
\date{\today}
\maketitle

\begin{abstract}
{We study the dimensional crossover from 1D to 2D in half-filled and lightly
doped, weakly interacting $N$-leg Hubbard ladders. In this case, the Hubbard
ladders are equivalent to a $N$-band model. Using renormalization group
techniques, we find, that the half-filled ladders exhibit (in the
spin-sector) an odd-even effect only below a crossover energy
$E_{c}\propto\exp[-\alpha\exp(\gamma N)]$ ($\alpha\ll 1$ and $\gamma\sim 1$
depend on the interaction strength and on the hopping matrix elements):
Below $E_{c}$, the dominant interactions take place within band pairs
$(j,N+1-j)$ [and within the band $(N+1)/2$ for $N$ odd], such that even-leg
ladders are an insulating spin-liquid, while odd-leg ladders have one gapless
spinon-mode. In contrast, above the energy-scale $E_{c}$, all bands are
interacting with each other and the system is a 2D-like (insulating)
antiferromagnet; we obtain an analytical expression for the Hamiltonian which
is similar to the 2D Heisenberg antiferromagnet. Bosonization techniques show,
that in the charge-sector the Mott insulator is as well below and above $E_{c}$
of the same type as in $N/2$ half-filled two-leg ladders. Doping away from
half-filling, we find that the effect of an increasing doping $\delta$ is very
similar to decreasing the number of legs~$N$: In both cases interactions between
unpaired bands are suppressed and thereby the antiferromagnetic correlations
reduced. The resulting band pairs form then insulating spin-liquids and when
doped, there is a spin-gap, but phase coherence exists only within the band pairs.
At higher doping levels $\delta_{c}=\delta_{c}(N)$, phase coherence between all
band pairs sets in and the system becomes a 2D-like $d$-wave superconductor
[$\delta_{c}(N)\rightarrow 0$ for $N\rightarrow\infty$].}
\end{abstract}


\begin{multicols}{2}

\narrowtext

\section{Introduction}

The physical properties of (quasi-)one-dimensional electron systems are often very
different from their 2D counterpart, for example, in (spin-$1/2$) Heisenberg
antiferromagnets (AFM): The (quasi-1D) ladders exhibit in the groundstate an
odd-even effect, i.e., odd-leg ladders have one gapless spinon-mode, while even-leg
ladders have a spin-gap~\cite{bib:DagRice}. The 2D Heisenberg model however, has two
gapless magnon-modes and long-range order~\cite{bib:CHN,bib:EMan}. Similarly,
(quasi-)1D conductors are generally Luttinger liquids (LL)~\cite{bib:Lutt,bib:Haldane},
while 2D and 3D electron systems are (usually) Fermi liquids (FL).

For the AFM Heisenberg model, the dimensional crossover from 1D to 2D has been studied in
Ref.~\cite{bib:Chak1}. Via the correspondence between number of legs $N$ and temperature
$T$, $1/T\propto N$, the nonlinear sigma-model allows one in this case to calculate
various physical properties; in particular, the spin-gap present in even-leg ladders
vanishes exponentially as a function of the number $N$ of coupled spin-chains. Coupled
LLs~\cite{bib:SchulzRev1} and the dimensional crossover to 2D have been investigated by
many different authors~\cite{bib:CCM,bib:BBT,bib:Arr}. One of the reasons was the search
for a non-FL (i.e., possible LL-like) phase in 2D~\cite{bib:And}. However, no clear
evidence for such a phase has been found so far~\cite{bib:CCM,bib:BBT,bib:Arr,bib:AV}.

In the half-filled and lightly doped Hubbard model, there is a \emph{strong competition}
between Cooper, umklapp, and AFM processes. For an unbiased determination of the
dominant phase, it is therefore advantageous to use the \emph{renormalization group} (RG)
technique~\cite{bib:Shankar}. In the (weakly interacting) 2D case, such RG studies found
AFM at and close to half-filling and $d$-wave superconductivity for sufficient hole
doping~\cite{bib:ZS,bib:HaMe,bib:HSFR,bib:BBD}. The dimensional crossover from 1D to 2D
was only investigated \emph{away} from half-filling (without umklapp and AFM
processes)~\cite{bib:LBF}. The authors concluded that the system becomes a FL for
$N\rightarrow\infty$.

In a previous work, we have studied in detail the groundstate properties of the
half-filled and lightly doped $N=3$ and 4-leg Hubbard ladders in the case of weak
interactions~\cite{bib:Led1}. In contrast to the large $U$ case, where the
half-filled Hubbard model converges onto the Heisenberg model of spin-1/2 \cite{bib:AAI},
in the small $U$ case, the \emph{charge degrees} of freedom are still \emph{present},
which gives insight in the effect of doping away from half-filling.

Here, we analyze the crossover from 1D to 2D for the half-filled (and lightly
doped) ladders. We will show that depending on the energy-scale (respectively
length-scale), the physical properties are very different and that the phases
associated with these properties are separated by crossover energies which
are a function of $N$. In particular, the quasi-1D analog of the 2D phase is
present at higher energies, implying that the crossover between
\emph{different} phases can be studied for $N$ \emph{finite}. An advantage in
quasi-1D compared to the 2D case is the limited number of different
interactions. This allows us to obtain (some) \emph{analytical} results and to
disentangle the interplay between AFMs (two gapless magnon-modes), ISLs
(insulating spin-liquid, i.e., only short-range correlations), and
superconductivity. The 2D limit ($N=\infty$) can then be taken within the
\emph{same} phase.

For weak interactions, the Hubbard ladders are equivalent to a $N$-band model,
characterized by Fermi velocities $v_{j}$. At half-filling, due to nesting of
the Fermi surface (FS), the Fermi velocities of the bands $j$ and $N+1-j$ are
equal (to be called ``band pairs'' below), such that AFM and umklapp processes
take place. Using RG techniques, we find that the half-filled ladders exhibit
(in the spin-sector) an odd-even effect only below a crossover energy
$E_{c}\propto\exp[-\alpha\exp(\gamma N)]$ ($\alpha\ll 1$ and $\gamma\sim 1$),
but are 2D-like (insulating) AFMs above the scale $E_{c}$. The reason is, that
below $E_{c}$ only interactions within band pairs $(j,N+1-j)$ [and within the
band $(N+1)/2$ for $N$ odd] are present, while above $E_{c}$ all bands are
interacting with each other. We obtain an analytical expression for the
Hamiltonian which is similar to the 2D Heisenberg AFM. Analyzing the charge-sector
by bosonization techniques~\cite{bib:Haldane,bib:SchulzRev2,bib:Voit}, we find,
that the Mott insulator is of the same type as in $N/2$ half-filled two-leg
ladders. The scale $E_{c}$ decreases fast to zero as the number of chains $N$ goes
to infinity. The 2D limit is thus an insulating AFM, in agreement with what has
been found in studies of the weakly interacting 2D Hubbard
model~\cite{bib:ZS,bib:HaMe,bib:HSFR,bib:BBD}.

\begin{figure}[t]
  \centerline{
   \psfig{file=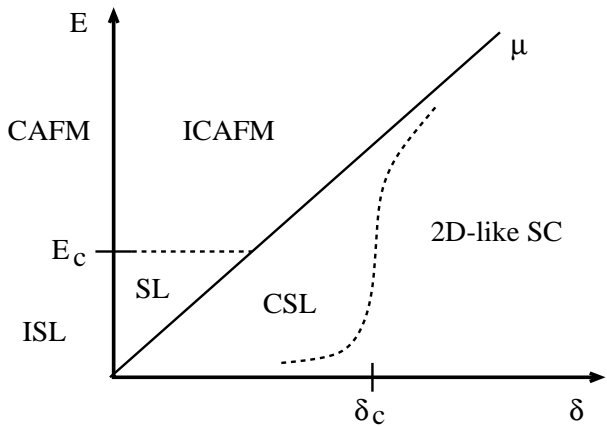,width=8cm,height=5.6cm}}
  \vspace{1mm}
  \caption{Schematic phase diagram for the (large) $N$-leg Hubbard ladders. At
  half-filling, the system below the energy-scale $E_{c}=E_{c}(N)$ is an insulating
  spin-liquid (ISL) and above $E_{c}$ an insulating, 2D-like, commensurate
  antiferromagnet (CAFM). Upon doping away from half-filling, the ladders become
  below the chemical potential $\mu$ (measured from its value at half-filling) a
  conducting spin-liquid (CSL), i.e., phase coherence is only present within pairs
  of Fermi momenta $(k_{F},\pi-k_{F})$ and $(\pi-k_{F},k_{F})$; above $\mu$, the
  antiferromagnetic correlations become incommensurate (ICAFM). Upon increasing
  doping, phase coherence between all Fermi momenta sets in and the system becomes
  a 2D-like $d$-wave superconductor.
  } 
  \label{f:sphd}
\end{figure}

Doping away from half-filling (at energies above $E_{c}$), we find that the effect of
a doping $\delta>0$ is the same as of a finite $N<\infty$: Doped holes suppress the
interactions between unpaired bands and thereby reduce the antiferromagnetic
correlations. The effect of the doping on the remaining interactions within a band
pair $(j,N+1-j)$ is then the same as when doping $N/2$ uncoupled ISLs; there is a
spin-gap and $d$-wave-like phase coherence between bands $j$ and $N+1-j$. At a higher
doping value $\delta_{c}=\delta_{c}(N)$, phase coherence between all band pairs sets
in and the $N$-leg ladders become a 2D-like $d$-wave superconductor
[$\delta_{c}(N)\rightarrow 0$ for $N\rightarrow\infty$]. The low-energy physics is
given by a $d$-wave BCS Hamiltonian. The origin of the superconducting instability is
the same as for the 2D case~\cite{bib:ZS,bib:HaMe,bib:HSFR}: A Kohn-Luttinger-type
attraction~\cite{bib:KoLu} is generated by AFM processes~\cite{bib:SLH}. Furthermore,
while at half-filling the antiferromagnetic correlations are commensurate, upon
doping, they become incommensurate. Fig.~\ref{f:sphd} gives a schematic overview.

Summarizing, the $N$-leg Hubbard model exhibits below the length-scale $\xi_{c}\propto 1/E_{c}$
almost the same physical properties as the 2D model (i.e., the same correlation functions
dominate), but in contrast to the 2D case, this quasi-1D analog can be treated (partially)
\emph{analytically}.

We do all our calculations for \emph{even}~$N$ and only refer to the odd $N$ case where
it is of interest. In Sec.~II, we introduce the $N$-leg ladder Hamiltonian, discuss the
various interactions, and specify along which ``path'' the dimensional crossover is
studied. In Sec.~III, we briefly revisit the results obtained previously in
Ref.~\cite{bib:Led1} and analyze the half-filled Hamiltonian as a function of the
energy-scale. We show that there exist for finite $N$ two clearly different phases,
separated by a crossover energy $E_{c}$. We derive an (analytical) expression for
$E_{c}$. In Sec.~IV, the phase above the energy $E_{c}$ is investigated in detail and
shown to be a 2D-like AFM. Finally, in Sec.~V, we treat the effect of doping away from
half-filling.

\section{$N$-leg Hubbard model}

The $N$-leg ($N$-chain) Hubbard model is given by $H=H_{0}+H_{\rm Int}$, where the kinetic
energy is
\begin{eqnarray}
  H_{0}&=&-t\sum_{x,i,s} d_{is}^{\dagger}(x+1)d_{is}(x)
    +{\rm H.c.}
  \nonumber\\
  &&-t_{\perp}\sum_{x,i,s}d_{i+1s}^{\dagger}(x)d_{is}(x)
    +{\rm H.c.}
\end{eqnarray}
and the interaction term is
\begin{equation}
  H_{\rm Int}=U\sum_{i,x}d_{i\uparrow}^{\dagger}(x)d_{i\uparrow}(x)
    d_{i\downarrow}^{\dagger}(x)d_{i\downarrow}(x).
\end{equation}
The hopping matrix elements along- and perpendicular to the chains are denoted
by $t$ and $t_{\perp}$ respectively and $U>0$ is the on-site repulsion. 
For our approach, it is advantageous first to diagonalize $H_{0}$,
\begin{equation}
  H_{0}=\sum_{j,s}\int dk\epsilon_{j}(k)\Psi_{js}^{\dagger}(k)\Psi_{js}(k),
\end{equation}
where $\Psi_{js}^{\dagger}$ and $\Psi_{js}$ are the creation- and annihilation
operators for the band~$j$. For open boundary conditions perpendicular to the
legs, the dispersion relations $\epsilon_{j}$ are given by ($j=1,\ldots,N$)
\begin{equation}
  \epsilon_{j}(k)=-2t\cos(k)-2t_{\perp}\cos\left[\pi j/(N+1)\right].
  \label{eq:DispRelN}
\end{equation}
The Fermi momenta in each band $k_{Fj}$ are determined by the chemical potential
$\mu=\epsilon_{j}(k_{Fj})$ and the filling $n=2(\pi N)^{-1}\sum k_{Fj}$. Since
we are only interested in the low-energy physics, we linearize the dispersion
relation at the FS, resulting in Fermi velocities $v_{j}=2t\sin(k_{Fj})$. We
introduce operators $\Psi_{R/Ljs}$ for right and left movers at the FS. At
half-filling, $\mu=0$, and
\begin{equation}
  v_{j}=v_{\jbar}=2\sqrt{t^{2}-t_{\perp}^{2}\cos[\pi j/(N+1)]^{2}},
\end{equation}
where $\jbar=N+1-j$. For $t=t_{\perp}$, $v_{1}=v_{N}\approx 2\pi t/N$, which
leads to a singular behavior for large $N$ (the quasi-1D analog of the van Hove
singularities in 2D). For the following, we take $t_{\perp}<t$ in order to
avoid singularities. Note that the FS is then not flat~\cite{bib:DAD}, but that
we still have nesting, i.e., $k_{Fj}+k_{F\jbar}=\pi$.

The crucial difference between quasi-1D and 2D are the interactions which control
the low-energy physics. The system is quasi-1D and only a finite number of different
interactions play a role, provided the energy difference between two neighboring
bands is larger than the largest energy-scale of the system, i.e., $te^{-t/U}$.
Therefore, we investigate the crossover between \emph{different phases} (ISL and AFM)
in quasi-1D, $U\ll t/\ln N,t_{\perp}/\ln N$, as a function of the \emph{energy} (this
can partially be done analytically). The 2D limit, $\ln N\gg t/U$, is then taken
within the \emph{same} phase. In other words, along this particular ``path'', the
dimensional crossover from 1D to 2D can be treated in a controlled way.

Since we are interested in the small-$U$ low-energy physics, we have to take into
account only the processes, which are present down to arbitrarily small
energy-scales. For the half-filled case, this implies that only umklapp processes
where the momenta add up exactly to $\pi$, $k_{Fl}+k_{Fm}=\pi$ are relevant. This
condition is only fulfilled for the band pairs $(j,\jbar)$, $k_{Fj}+k_{F\jbar}=\pi$.
When doping away from half-filling, the chemical potential (measured from its value
at half-filling) introduces a low-energy cutoff for the various umklapp processes.
The interactions which have to be taken into account initially are then different.

At half-filling, the interacting part of the Hamiltonian consists of forward (f),
Cooper (c), and umklapp (u) processes between~2 different bands and (non-)umklapp
processes between~4 different bands (in our notation, we follow Refs.~\cite{bib:Led1,bib:LBFso8}),
$H_{\rm Int}=H^{2B}+H^{4B}$, where
\begin{eqnarray}
  H^{2B}&=&\sum_{i\neq j}\int dx\left(f_{ij}^{\rho}J_{Rii}J_{Ljj}
    -f_{ij}^{\sigma}{\bf J}_{Rii}\cdot{\bf J}_{Ljj}\right.
  \nonumber\\
  &&\left.+c_{ij}^{\rho}J_{Rij}J_{Lij}
    -c_{ij}^{\sigma}{\bf J}_{Rij}\cdot{\bf J}_{Lij}\right)
  \nonumber\\
  &&+\sum_{j}\int dx\left(c_{jj}^{\rho}J_{Rjj}J_{Ljj}
    -c_{jj}^{\sigma}{\bf J}_{Rjj}\cdot{\bf J}_{Ljj}\right)
  \nonumber\\
  &&+\sum_{j}\int dx\left[\left(u_{j\jbar\jbar j}^{\rho}I_{Rj\jbar}^{\dagger}I_{L\jbar j}
  -u_{j\jbar\jbar j}^{\sigma}{\bf I}_{Rj\jbar}^{\dagger}\cdot{\bf I}_{L\jbar j}\right.\right.
  \nonumber\\
  &&\left.\left.
  +u_{jj\jbar\jbar}^{\rho}I_{Rjj}^{\dagger}I_{L\jbar\jbar}\right)+{\mathrm H.c.}\right]
\end{eqnarray}
where the U(1) and SU(2) currents are defined as
\begin{equation}
  J_{hij}=\sum_{s}\Psi_{his}^{\dagger}\Psi_{hjs}
  \label{eq:ChargeCurrents}
\end{equation}
and
\begin{equation}
  J_{hij}^{p}=\frac{1}{2}\sum_{s,s^{\prime}}\Psi_{his}^{\dagger}
    \tau_{ss^{\prime}}^{p}\Psi_{hjs^{\prime}},
  \label{eq:SpinCurrents}
\end{equation}
where again $h=R/L$ and the $\tau^{p}$ are the Pauli matrices ($p=x,y,z$). Due to symmetry,
$f_{ij}^{\rho,\sigma}=f_{ji}^{\rho,\sigma}$ and $c_{ij}^{\rho,\sigma}=c_{ji}^{\rho,\sigma}$.
The umklapp currents are
\begin{equation}
  I_{hij}=\sum_{s,s^{\prime}}\Psi_{his}\epsilon_{ss^{\prime}}\Psi_{hjs^{\prime}}
\end{equation}
and
\begin{equation}
  I_{hij}^{p}=\frac{1}{2}\sum_{s,s^{\prime}}
    \Psi_{his}\left(\epsilon\tau^{p}\right)_{ss^{\prime}}\Psi_{hjs^{\prime}},
\end{equation}
where $\epsilon=-i\tau^{y}$. The 4-band interactions involve processes between
two different band pairs $(j,\jbar)$ and $(k,\kbar)$,
\begin{equation}
  H^{4B}=\sum_{j=1}^{N/2-1}\sum_{k=j+1}^{N-j}\int dx{\mathcal H}_{jk}^{4B},
\end{equation}
where
\begin{eqnarray}
  {\mathcal H}_{jk}^{4B}=c_{jk\kbar\jbar}^{\rho}\left(J_{Rjk}J_{L\kbar\jbar}
    +J_{R\jbar\kbar}J_{Lkj}+{\mathrm H.c.}\right)
  \nonumber\\  
    -c_{jk\kbar\jbar}^{\sigma}\left({\mathbf J}_{Rjk}\cdot
    {\mathbf J}_{L\kbar\jbar}+{\mathbf J}_{R\jbar\kbar}\cdot
    {\mathbf J}_{Lkj}+{\mathrm H.c.}\right)
  \nonumber\\
    +u_{jk\kbar\jbar}^{\rho}\left(I_{Rjk}^{\dagger}I_{L\kbar\jbar}
    +I_{R\kbar\jbar}^{\dagger}I_{Ljk}+{\mathrm H.c.}\right)
  \nonumber\\
    +u_{j\jbar k\kbar}^{\rho}\left(I_{Rj\jbar}^{\dagger}I_{Lk\kbar}
    +I_{Rk\kbar}^{\dagger}I_{Lj\jbar}+{\mathrm H.c.}\right)
  \nonumber\\
    -u_{jk\kbar\jbar}^{\sigma}\left({\mathbf I}_{Rjk}^{\dagger}\cdot
    {\mathbf I}_{L\kbar\jbar}+{\mathbf I}_{R\kbar\jbar}^{\dagger}
    \cdot{\mathbf I}_{Ljk}+{\mathrm H.c.}\right).
  \label{eq:Hcu4B}
\end{eqnarray}
In our case $v_{j}=v_{\jbar}$ implying
$c_{jk\kbar\jbar}^{\rho,\sigma}=c_{j\kbar k\jbar}^{\rho,\sigma}$,
$u_{jk\kbar\jbar}^{\rho,\sigma}=u_{j\kbar k\jbar}^{\rho,\sigma}$.
Note that the 4-band couplings describe \emph{AFM processes}, i.e., the
terms $\Psi_{Rjs}^{\dagger}\Psi_{L\jbar s^{\prime}}$ result in a momentum transfer
of $\pi$ (along the chains).

\section{Dimensional crossover}

Using the RG method, we study the half-filled $N$-leg ladders as a function of the \emph{energy-scale}.
We show that there exist two clearly different phases, separated by a crossover energy
$E_{c}\ll t$. The phase below $E_{c}$ is an ISL, i.e., 4-band-AFM processes are suppressed
and the physics is dominated by 2-band umklapp and Cooper processes~\cite{bib:Led1}. In
contrast, in the phase above $E_{c}$ (but still below the bandwidth $t$) the 4-band-AFM
processes are large and dominating (this phase is investigated in detail in Sec.~IV). First
(Sec.~III~A), we revisit the results for $N$-leg ladders obtained previously in
Ref.~\cite{bib:Led1} and give a criterion for the couplings allowing for a (rather) precise
distinction between the two phases. In Sec.~III~B, we calculate the crossover energy
$E_{c}$ as a function of $N$.

\subsection{ISL phase in even-leg ladders}

In Ref.~\cite{bib:Led1}, we have shown, that at half-filling (for $N$ even)
the band pair $(j,\jbar)$ scales towards an ISL and is frozen out at the
energy-scale $te^{-\alpha_{j}v_{j}/U}$, where $\alpha_{j}\sim 1$. The
low-energy Hamiltonian is then the same as of $N/2$ half-filled two-leg
ladders. For increasing $N$, the initial-value $U/t$ (and therefore the
energy-scale), which leads to this decoupling (in particular to a suppression
of the 4-band-AFM processes) and an ISL phase, decreases fast. Next, using
bosonization techniques~\cite{bib:Haldane,bib:SchulzRev2,bib:Voit}, we show
that the RG flow of the \emph{Luttinger liquid parameter} for spin triplet
pairing, $K_{\sigma j+}$, allows one to determine the energy-scale, where the
spin-gap opens. This energy-scale can then be interpreted as the crossover
energy between the ISL phase at low energies and the AFM phase at higher
energies (which has no spin-gap).

Bosonization is the method of rewriting fermionic creation (annihilation)
operators in terms of field operators $\Phi_{\alpha}$ and $\Pi_{\alpha}$
satisfying the commutation relation
$[\Phi_{\alpha}(x),\Pi_{\alpha}(y)]=i\delta(x-y)$ ($\alpha$ labels the charge and
spin-modes for the different bands). The dual field of $\Phi_{\alpha}$ is defined
as $\partial_{x}\theta_{\alpha}=\Pi_{\alpha}$. The fields $\Phi_{\alpha}$ contain
the (physical) charge respectively spin of the particle, i.e., the
charge/spin-density is $\rho_{\alpha}=\partial_{x}\Phi_{\alpha}$, while the fields
$\theta_{\alpha}$ can be interpreted as the phase of the charge-part respectively
the spin-part (for ladders, see, e.g., Refs.~\cite{bib:LBF,bib:Led1,bib:LBFso8}).
We introduce the fields
\begin{equation}
  \Phi_{\nu j\jbar\pm}=\frac{1}{\sqrt{2}}\left(\Phi_{\nu j}\pm\Phi_{\nu\jbar}\right)
\end{equation}
and
\begin{equation}
  \Pi_{\nu j\jbar\pm}=\frac{1}{\sqrt{2}}\left(\Pi_{\nu j}\pm\Pi_{\nu\jbar}\right)
\end{equation}
combining the fields of the bands $j$ and $\jbar$ ($\nu=\rho,\sigma$), e.g.,
the spin-fields $\Phi_{\sigma j\jbar\pm}$ describe triplet/singlet pairs
between the momenta $k_{Fj}$ and $k_{F\jbar}$.

The bosonized Hamiltonian takes the following form. The noninteracting
Hamiltonian $H_{0}$ plus the contributions of the charge-currents
$f_{j\jbar}^{\rho}$ and $c_{jj}^{\rho}$ and the $z$-component of the
spin-currents $f_{j\jbar}^{\sigma}$ and $c_{jj}^{\sigma}$ can be written as
\begin{equation}
  \tilde{H}_{0}=\sum_{j,\nu,\mu}\int dx \frac{v_{\nu j\mu}}{2}\left[
      \frac{1}{K_{\nu j\mu}}(\partial_{x}\Phi_{\nu j\jbar\mu})^{2}
      +K_{\nu j\mu}\Pi_{\nu j\jbar\mu}^{2}\right],
  \label{eq:H0Bos}
\end{equation}
where $\nu=\rho,\sigma$ and $\mu=\pm$ and the Luttinger liquid parameters (LLP) are
\begin{equation}
  K_{\rho j\pm}=\sqrt{\frac{\pi v_{j}-(c_{jj}^{\rho}\pm f_{j\jbar}^{\rho})}
                           {\pi v_{j}+(c_{jj}^{\rho}\pm f_{j\jbar}^{\rho})}}
\end{equation}
and
\begin{equation}
  K_{\sigma j\pm}=\sqrt{\frac{4\pi v_{j}+(c_{jj}^{\sigma}\pm f_{j\jbar}^{\sigma})}
                             {4\pi v_{j}-(c_{jj}^{\sigma}\pm f_{j\jbar}^{\sigma})}}.
\end{equation}
The velocities are $v_{\nu j\mu}\sim v_{j}$.
The interacting part of the $N$-leg ladder Hamiltonian contains (apart form other
terms) the expression~\cite{bib:LBFso8}
\begin{equation}
  -\cos\left(\sqrt{4\pi}\Phi_{\sigma j\jbar+}\right)\times{\rm other\;cosine\;terms}.
\end{equation}
Fields which appear in a cosine become ``pinned'' in order to minimize the energy,
e.g., the term $-\cos(\sqrt{4\pi}\Phi_{\sigma j\jbar+})$ implies (quasiclassicaly)
$\Phi_{\sigma j\jbar+}\approx 0$ and the corresponding spin-mode acquires a \emph{gap}.

The LLPs determine the exponents of correlation functions~\cite{bib:SchulzRev2,bib:Voit}
and --- together with cosine-terms --- the pinned fields. Here, as we show in the following,
the flow of the LLP $K_{\sigma j+}$ can be used to determine the energy-scale where the
spin-gap in the $\sigma+$ sector opens.
Note that initially (i.e., at high energies) $K_{\sigma j+}>1$ and $K_{\sigma j-}=1$
(and $K_{\rho j+}<1$, $K_{\rho j-}=1$). For the $N$-leg ladders, the RG flow of the
couplings is such that $c_{jj}^{\sigma}\approx -t$ and $f_{j\jbar}^{\sigma}\approx 0$
at low energies, implying $K_{\sigma j\pm}<1$. Hence, while $K_{\sigma j-}$ is always
$\leq 1$, there is an energy, where $K_{\sigma j+}$ becomes smaller than 1, i.e.,
$r_{j}=c_{jj}^{\sigma}+f_{j\jbar}^{\sigma}\leq 0$.
The canonical transformation
\begin{equation}
  \tilde{\Phi}_{\sigma j\jbar+}=\Phi_{\sigma j\jbar+}/\sqrt{K_{\sigma j+}},\;\;\;
  \tilde{\theta}_{\sigma j\jbar+}=\theta_{\sigma j\jbar+}\sqrt{K_{\sigma j+}}
\end{equation}
eliminates $K_{\sigma j+}$ from $\tilde{H}_{0}$ (it just renormalizes the velocities)
and results in
\begin{equation}
  -\cos\left(\sqrt{4\pi K_{\sigma j+}}\tilde{\Phi}_{\sigma j\jbar+}\right)
    \times{\rm other\;cosine\;terms}.
\end{equation}
A large $K_{\sigma j+}$ therefore tends to depin the field $\tilde{\Phi}_{\sigma j\jbar+}$,
while a small $K_{\sigma j+}$ results in a strong pinning of the field
$\tilde{\Phi}_{\sigma j\jbar+}$ (for the dual field $\tilde{\theta}_{\sigma j\jbar+}$, it
is vice versa). The crossover between the two regimes takes place at $K_{\sigma j+}=1$
(since we are at finite energy, it is a crossover and not a transition). The ISL phase
therefore requires $r_{j}<0$ ($K_{\sigma j+}<1$) and $r_{j}$ can be used in the RG flow to
determine a crossover energy. We note that for the pure sine-Gordon model (at zero
temperature), the value $K_{\alpha}=1$ rigorously separates the massive from the massless
phase~\cite{bib:Gogolin}.

\begin{figure}[t]
  \centerline{
   \psfig{file=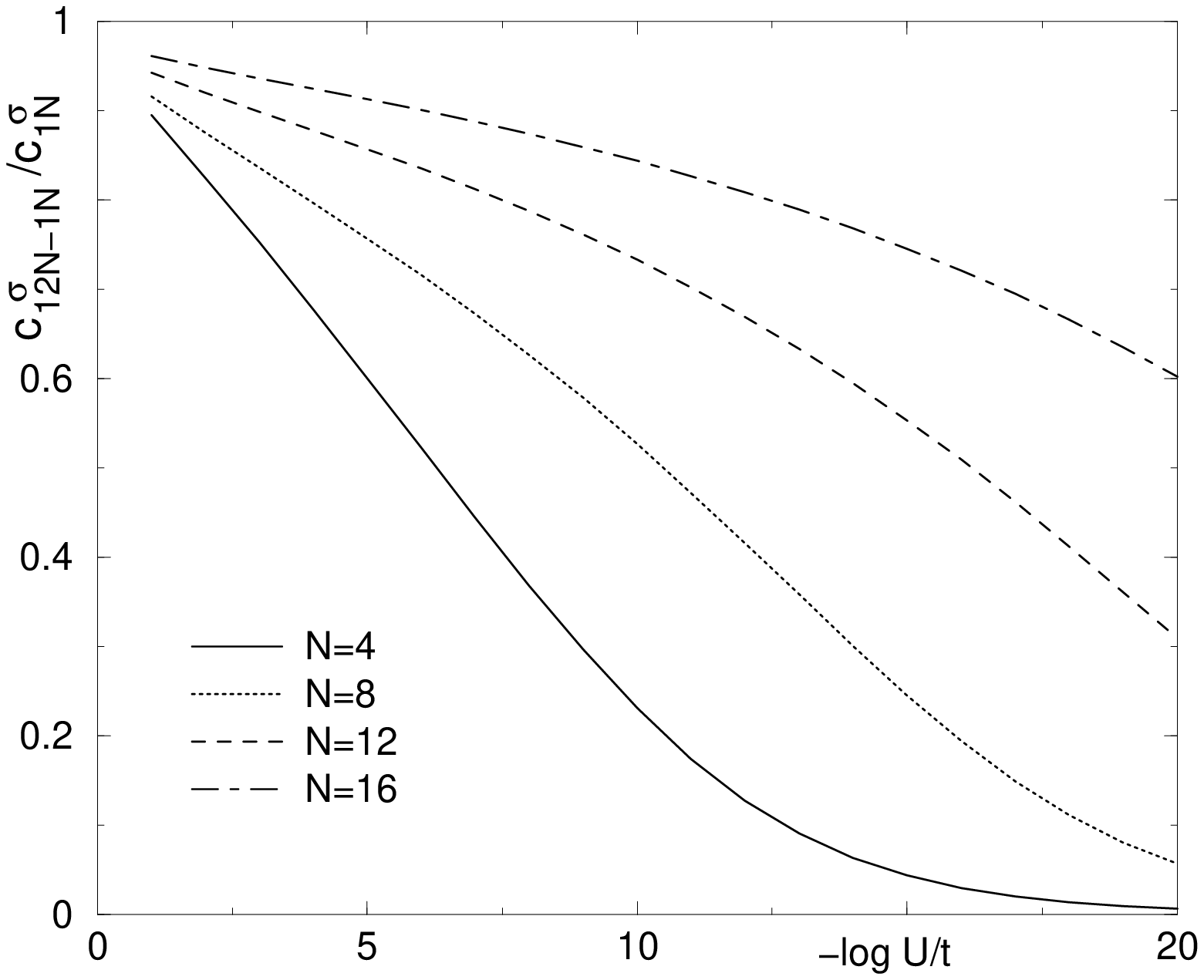,width=8cm,height=6.7cm}}
  \centerline{
   \psfig{file=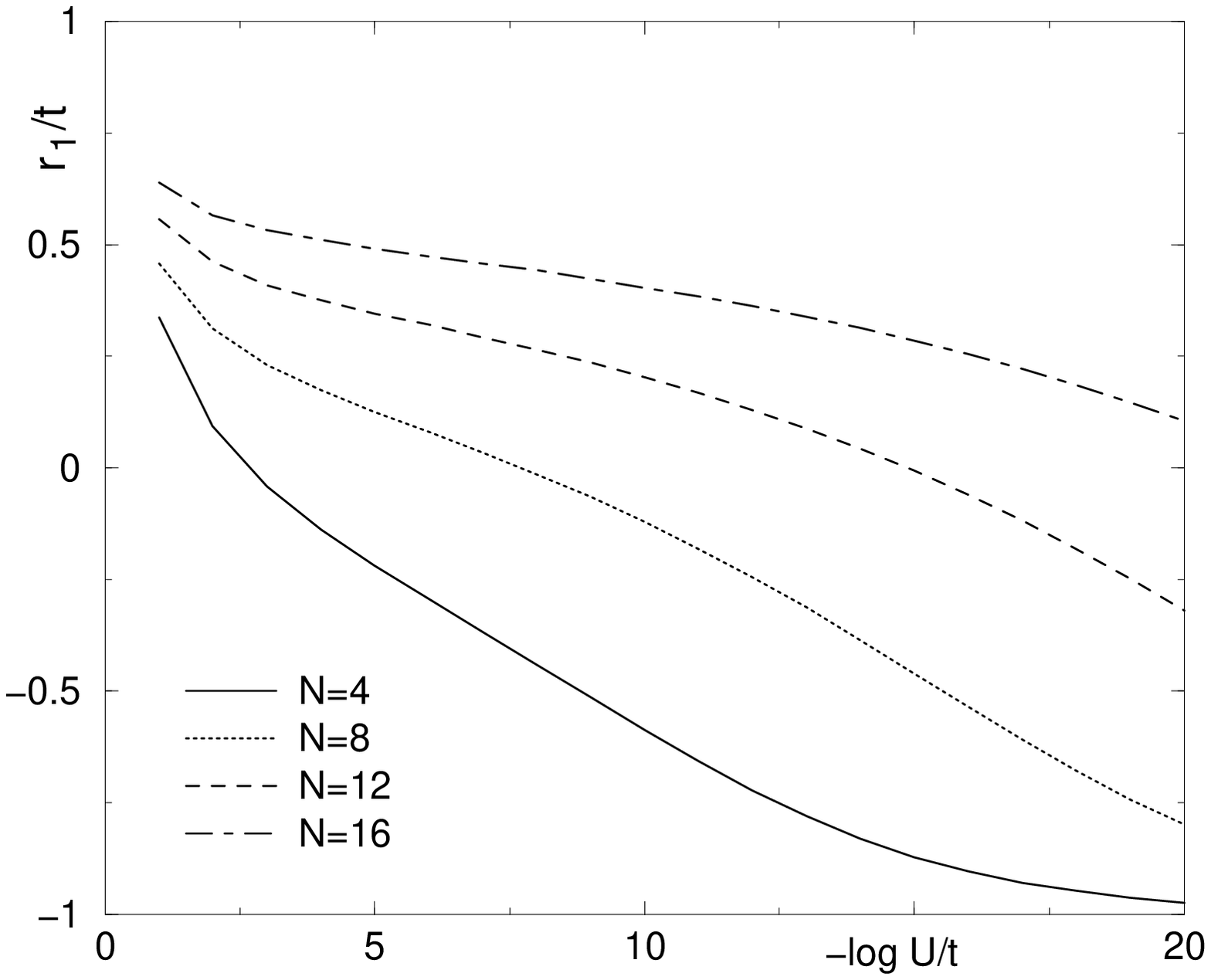,width=8cm,height=6.7cm}}
  \vspace{1mm}
  \caption{The upper figure shows the strong-coupling value of the ratio of a
  4-band-AFM and a 2-band Cooper coupling, $c_{12N-1N}^{\sigma}/c_{1N}^{\sigma}$,
  as a function of the initial-value $U/t$.
  For increasing $N$, the 4-band-AFM couplings are only suppressed for very small
  initial-values. Similarly, the lower figure displays the strong-coupling value
  of the sum of couplings $r_{1}=c_{11}^{\sigma}+f_{1N}^{\sigma}$. 
  The initial-values which lead to $r_{1}<0$ (i.e., an ISL) become very small as
  $N$ becomes large. We take both times $t_{\perp}/t=0.95$.
  } 
  \label{f:CFlow}
\end{figure}

\subsection{Determination of the crossover energy}

Performing a numerical integration of the full set of RG equations (RGEs)~\cite{bib:Led2}, we
calculate the crossover energy $E_{c}=E_{c}(N)$, which separates ladder-like from
2D-AFM-like behavior (the RGEs are a generalization of the 3-leg RGEs given
in Ref.~\cite{bib:Led1}; the RGEs relevant for the AFM phase are given below).

In Fig.~\ref{f:CFlow}, we have plotted the strong-coupling value of the ratio of
a 4-band AFM and a 2-band Cooper coupling, $c_{12N-1N}^{\sigma}/c_{1N}^{\sigma}$,
and of $r_{1}=c_{11}^{\sigma}+f_{1N}^{\sigma}$ for $N=4,8,12,16$ (we fix the
strong-coupling value of $c_{1N}^{\sigma}$ at the bandwidth $t$).
We take $t_{\perp}/t=0.95$, because for $t_{\perp}/t=1$, the band pair $(1,N)$ is close
to the van Hove singularities, such that its behavior would no more be typical for the
system. When the number of chains, $N$, increases, the
initial-value $U/t$ which leads to a suppression of the 4-band-AFM couplings
decreases fast. Similarly, for large $N$, the $r_{1}$ becomes only negative
when $U/t$ is very small. Note that for the other band pairs $(j,\jbar)$,
the corresponding energy-scales are \emph{lower}, since the velocities $v_{j}$
are larger and the gaps are of the order of $t\exp(-\alpha_{j}v_{j}/U)$.

\begin{figure}[t]
  \centerline{
   \psfig{file=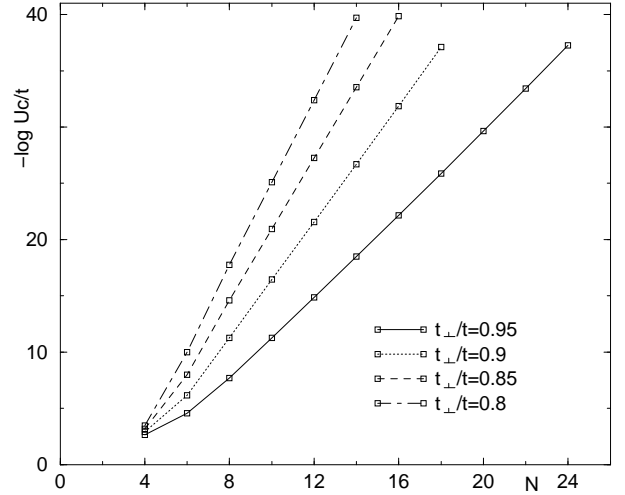,width=8cm,height=6.7cm}}
  \vspace{1mm}
  \caption{The sum of couplings $r_{j}=c_{jj}^{\sigma}+f_{j\jbar}^{\sigma}$ allows
    to determine between ISL and AFM: For $r_{j}<0$ the band pair $(j,\jbar)$ is an ISL and
    for $r_{j}>0$ an AFM. We have plotted the initial-value $U_{c}/t$, which leads to
    $r_{1}\approx 0$ at the energy-scale, where the first coupling of the band pair $(1,N)$
    has grown up to the bandwidth $t$. The figure shows, that $-\log U_{c}(N)/t$ is
    in good approximation a linear function of $N$
    (the squares are the calculated initial-values and the lines are a guide to the eye).
    } 
  \label{f:UCross}
\end{figure}

Hence, both the 4-band-AFM processes and the $r_{j}$
have a strong and similar dependence on the energy-scale. For small enough initial-values
$U/t$, the \emph{differences} between the Fermi velocities lead
to a decoupling into band pairs. However, since $v_{j}-v_{j+1}\propto 1/N$, for increasing
$N$, this decoupling becomes suppressed and the physics is dominated by 4-band-AFM processes.

We calculate the crossover energy $E_{c}$ as follows. We determine initial-values
$U_{c}=U_{c}(N)$ such that $r_{1}\approx 0$ at the scale where the first coupling
of the band pair $(1,N)$ has grown up to the bandwidth $t$. For initial-values
$U<U_{c}(N)$, a spin-gap then opens, starting in the band pair $(1,N)$,
and the 4-band-AFM processes become suppressed. We find that $U_{c}(N)$ can be
fitted by $U_{c}(0)\exp(-\gamma N)$, where $\gamma\sim 1$ (for
$t_{\perp}/t=0.8,0.85,0.9,0.95$, we obtain $\gamma=3.7,3.2,2.6,1.9$)
and $U_{c}(0)\gg t$, see Fig.~\ref{f:UCross}. On the other hand, an initial-value
$U$ corresponds to an energy-scale (gap) $t\exp(-t/U)$. The crossover energy $E_{c}$ is
therefore given by
\begin{equation}
  E_{c}\sim t\exp[-\alpha\exp(\gamma N)],
\end{equation}
where $\alpha\sim t/U_{c}(0)\ll 1$. Note that the energy $E_{c}$ is also the upper limit
for the spin-gap in the ISL phase. For large $U$ (Heisenberg AFM), the corresponding
scale (spin-gap) is $J\exp(-0.68\,N)$, where $J=4t^{2}/U$ (see also table~I)~\cite{bib:Chak1}.
While for large $U$ the spin-gap decreases exponentially, in the small-$U$ case the
decrease is \emph{double-exponentially}.

As a result, the ``phase'' of the 2D system is for finite $N$ \emph{present} above an
energy-scale $E_{c}=E_{c}(N)$, where $E_{c}(N)\rightarrow 0$ for $N\rightarrow\infty$.
Equivalently, below the length-scale $\xi_{c}\propto 1/E_{c}$,
the dominating correlation functions are the same as in the 2D system.

\section{The 2D-like AFM phase}

First (Sec.~IV~A), using the RG method, we calculate the asymptotic ratios
of the couplings and the charge-gap of the phase present above $E_{c}$.
Comparing our results with the flow well away from half-filling~\cite{bib:LBF},
we conclude, that AFM processes (at high energies) are necessary for an
instability in the RG flow of the $N$-leg Hubbard model. In Sec.~IV~B, we show
that the phase above $E_{c}$ is the quasi-1D counterpart of a 2D insulating
AFM. In particular, there is no odd-even effect and the interacting part of
the Hamiltonian is similar to the 2D Heisenberg AFM. Using
bosonization techniques, we obtain, that the system is Mott insulating and that
the insulator is of the same type as in the half-filled two-leg ladder, see
Sec.~IV~C. For the spin-sector, we find the same physical properties as for the
Heisenberg AFM, i.e., two gapless magnon modes, spinon confinement, and
long-range order in the 2D limit.

\subsection{Asymptotic ratios of the couplings}

We give the RGEs of the 4-band-AFM couplings for large $N$, i.e., keeping in the
RGEs only sums over $N$ products of couplings and dropping the (2-band)
contributions of the order of~1. We then determine (analytically) the asymptotic
ratios of the couplings for Hubbard initial-values and the size of the charge-gap.

\subsubsection{RGEs}

The RGEs of the 4-band interactions are given by a sum
over $N/2$ products of 4-band couplings
\begin{eqnarray}
  \frac{dc_{jk\kbar\jbar}^{\rho}}{dl}&=&\sum_{i\neq j,k}^{N/2}\frac{1}{v_{i}}
    \left(c_{ji\ibar\jbar}^{\rho}c_{ki\ibar\kbar}^{\rho}
    +u_{ji\ibar\jbar}^{\rho}u_{ki\ibar\kbar}^{\rho}\right.
  \nonumber\\
    &&\left.+\frac{3}{16}c_{ji\ibar\jbar}^{\sigma}c_{ki\ibar\kbar}^{\sigma}
    +\frac{3}{16}u_{ji\ibar\jbar}^{\sigma}u_{ki\ibar\kbar}^{\sigma}\right)
  \nonumber\\
  \frac{du_{jk\kbar\jbar}^{\rho}}{dl}&=&\sum_{i\neq j,k}^{N/2}\frac{1}{v_{i}}
     \left(c_{ji\ibar\jbar}^{\rho}u_{ki\ibar\kbar}^{\rho}
     -\frac{3}{16}c_{ji\ibar\jbar}^{\sigma}u_{ki\ibar\kbar}^{\sigma}
     +k\leftrightarrow j\right)
  \nonumber\\
  \frac{dc_{jk\kbar\jbar}^{\sigma}}{dl}&=&\sum_{i\neq j,k}^{N/2}\frac{1}{v_{i}}
     \left(-\frac{1}{2}c_{ji\ibar\jbar}^{\sigma}c_{ki\ibar\kbar}^{\sigma}
                    +c_{ji\ibar\jbar}^{\sigma}c_{ki\ibar\kbar}^{\rho}
                    +c_{ki\ibar\kbar}^{\sigma}c_{ji\ibar\jbar}^{\rho}\right.
  \nonumber\\
                    &&\left.-\frac{1}{2}u_{ji\ibar\jbar}^{\sigma}u_{ki\ibar\kbar}^{\sigma}
                    -u_{ji\ibar\jbar}^{\sigma}u_{ki\ibar\kbar}^{\rho}
                    -u_{ki\ibar\kbar}^{\sigma}u_{ji\ibar\jbar}^{\rho}\right)
  \nonumber\\
  \frac{du_{jk\kbar\jbar}^{\sigma}}{dl}&=&\sum_{i\neq j,k}^{N/2}\frac{1}{v_{i}}
      \left(-\frac{1}{2}c_{ji\ibar\jbar}^{\sigma}u_{ki\ibar\kbar}^{\sigma}
      +u_{ji\ibar\jbar}^{\sigma}c_{ki\ibar\kbar}^{\rho}\right.
  \nonumber\\
      &&\left.-u_{ji\ibar\jbar}^{\rho}c_{ki\ibar\kbar}^{\sigma}+k\leftrightarrow j\right),
\end{eqnarray}
where the energy-scale is related to $l$ by $E\sim te^{-\pi l}$. Defining
\begin{equation}
  h_{jk}^{\pm}=\frac{c_{jk\kbar\jbar}^{\sigma}}{4}+\frac{u_{jk\kbar\jbar}^{\sigma}}{4}
               \pm\left(c_{jk\kbar\jbar}^{\rho}-u_{jk\kbar\jbar}^{\rho}\right)
\end{equation}
we find from the above RGEs, that
\begin{equation}
  \frac{d}{dl}h_{jk}^{+}=\sum_{i\neq j,k}^{N/2}\frac{1}{v_{i}}h_{ji}^{+}h_{ki}^{+}.
\end{equation}
Given the Hubbard initial-values
$4c_{jk\kbar\jbar}^{\rho}=c_{jk\kbar\jbar}^{\sigma}=2u_{jk\kbar\jbar}^{\rho}=2U/(N+1)$ and
$u_{jk\kbar\jbar}^{\sigma}=0$, we obtain $h_{jk}^{+}(l)=0$ for all $l$ (this fixed point
is stabilized by the 2-band interactions). Since
\begin{equation}
  \frac{d}{dl}h_{jk}^{-}=-\sum_{i\neq j,k}\frac{1}{v_{i}}\left[h_{ji}^{-}h_{ki}^{-}
                          +(h_{ji}^{+}+h_{ji}^{-})(h_{ki}^{+}+h_{ki}^{-})\right]
\end{equation}
and $h_{jk}^{-}(0)>0$, the $h_{jk}^{-}$ flow to 0, $h_{jk}^{-}(l)\rightarrow 0$
for increasing $l$. From the flow of the $h_{jk}^{\pm}$, we then calculate the
flow of the 4-band couplings and obtain the asymptotic ratios
\begin{equation}
  t\sim 3g_{jk}=4c_{jk\kbar\jbar}^{\rho}=3c_{jk\kbar\jbar}^{\sigma}
        =4u_{jk\kbar\jbar}^{\rho}=-3u_{jk\kbar\jbar}^{\sigma}.
  \label{eq:Ratios4B}
\end{equation}

\begin{table}
\begin{center}
\caption{The table shows the correspondence between crossover energy (spin-gap)
and energy-scale of the 2D system for small and large $U$. The exponentially small
charge-gap for small $U$ leads to a double exponentially suppressed 1D-2D crossover
energy (the result for the spin-gap in the large $U$ case is from Ref.~[6]).}
\vspace{3mm}
\begin{tabular}{l c c c c}
     &  2D energy-scale & crossover energy \\
\hline
$U\ll t$ & $t\exp(-\lambda t/U)$ & $t\exp[-\alpha\exp(\gamma N)]$ \\
$U\gg t$ & $J$                   & $J\exp(-0.68\,N)$ \\
\end{tabular}
\end{center}
\end{table}

Note that $u_{j\jbar k\kbar}^{\rho}$ remains always small. The 2-band and single-band
interactions are for large $N$ dominantly renormalized by the 4-band interactions.
The asymptotic ratios of the 2-band and single-band couplings can therefore be calculated
by inserting the ratios~(\ref{eq:Ratios4B}) in the RGEs for these couplings~\cite{bib:Led1}.
We then find, that the following couplings of the band pairs $(j,\jbar)$ grow and approach
fixed ratios:
\begin{eqnarray}
  t\sim 3g_{j}&=&4f_{j\jbar}^{\rho}=3f_{j\jbar}^{\sigma}=4c_{j\jbar}^{\rho}=3c_{j\jbar}^{\sigma}
  \nonumber\\
             &=&4u_{j\jbar\jbar j}^{\rho}=8u_{jj\jbar\jbar}^{\rho}=3u_{j\jbar\jbar j}^{\sigma}.
  \label{eq:Ratios2B}
\end{eqnarray}
The other 2-band and single-band couplings stay small; in particular the single-band
SU(2) processes are small, $|c_{jj}^{\sigma}|\ll t$, but they have become \emph{attractive},
$c_{jj}^{\sigma}<0$.

For bands $k$ and $j$, which are close together on the FS, $k\rightarrow j$, the 4-band
coupling becomes the same as the corresponding 2-band couplings,
$g_{jk}\approx g_{j}\approx g_{k}$. In the limit $N\rightarrow\infty$, all the (2 and
4-band) $g$-couplings take the same value, $g_{j}=g_{m}=g_{kl}$. The gap
$\Delta$ is then similarly calculated as the asymptotic ratios; using
that
\begin{equation}
  \frac{d}{dl}s_{jk}=\sum_{i\neq j,k}^{N/2}\frac{1}{v_{i}}s_{ji}s_{ki},
\end{equation}
where
\begin{equation}
  s_{jk}=\frac{c_{jk\kbar\jbar}^{\sigma}}{4}-\frac{u_{jk\kbar\jbar}^{\sigma}}{4}
               +c_{jk\kbar\jbar}^{\rho}+u_{jk\kbar\jbar}^{\rho},
\end{equation}
we find for the scale of divergence
\begin{equation}
  l_{c}=\frac{N+1}{U\sum_{j}^{N} 1/v_{j}}.
  \label{eq:DivSc}
\end{equation}
In particular, $l_{c}=2t/U$ for $t_{\perp}/t=0$ and
\begin{equation}
  l_{c}=\frac{\pi t}{U}\frac{1}{\ln(2N/\pi)}
  \label{eq:VHS}
\end{equation}
for $t=t_{\perp}$, where $\ln N<t/U$ for the validity of our calculations. The logarithmic
corrections come from the fact that $v_{1}=v_{N}\sim t/N$ (van Hove singularities).
The gap becomes
\begin{equation}
  \Delta\sim te^{-l_{c}}=te^{-\lambda t/U},
  \label{eq:2DGap}
\end{equation}
where $\lambda$ is a function of  $t_{\perp}/t$ and of the order of~1, $\lambda\sim 1$
(see also table~I).

\subsubsection{Discussion}

It is instructive to compare our result with the situation well away from half-filling,
where the umklapp interactions can be neglected. In Ref.~\cite{bib:LBF}, it has been
shown, that without umklapp interactions and for repulsive interactions, $U>0$, the
large $N$ limit is a FL, i.e., the typical gap-size scales exponentially to $0$,
$\propto e^{-N}$. The reason is, that the forward scattering (f) processes, which tend
to drive the system towards a RG instability give contributions of the order of~1, while
the Cooper processes (c), which drive the system towards a FL, have a weight $\propto N$.

At half-filling, the situation is completely different: Here, the 4-band AFM processes
with a weight $\propto N$ are (entirely) responsible for the RG instability at the
energy-scale $\Delta\sim te^{-\lambda t/U}$. Therefore, the $N$-leg Hubbard model
exhibits for large $N$ only a RG instability, when 4-band AFM processes are (at least
initially) present. The van Hove singularities, see Eq.~(\ref{eq:VHS}), lead to a further
increase of the gap-size (but are not the reason for the instability).

\subsection{Effective low-energy Hamiltonian}

We show that above $E_{c}$, the Hamiltonian of the half-filled $N$-leg Hubbard ladders is
similar to the 2D Heisenberg AFM.

At low energies (but above $E_{c}$), the ratios of the couplings are given
by Eqs.~(\ref{eq:Ratios4B}) and (\ref{eq:Ratios2B}), such that the U(1) and
SU(2) 4-band interactions, see Eq.~(\ref{eq:Hcu4B}), simplify to
\begin{eqnarray}
  3J_{Rjk}J_{L\kbar\jbar}-4{\mathbf J}_{Rjk}\cdot{\mathbf J}_{L\kbar\jbar}
    =2\Psi_{Rjs}^{\dagger}\Psi_{L\jbar s}\Psi_{L\kbar\bar{s}}^{\dagger}\Psi_{Rk\bar{s}}
  \nonumber\\
  -2\Psi_{Rjs}^{\dagger}\Psi_{L\jbar s}\Psi_{L\kbar s}^{\dagger}\Psi_{Rks}
    -4\Psi_{Rjs}^{\dagger}\Psi_{L\jbar\bar{s}}\Psi_{L\kbar\bar{s}}^{\dagger}\Psi_{Rks}
  \label{eq:AF4Bnu}
\end{eqnarray}
(for $s=\uparrow$, $\bar{s}=\downarrow$ and vice versa) and
\begin{eqnarray}
  3I_{Rjk}^{\dagger}I_{L\kbar\jbar}+4{\mathbf I}_{Rjk}^{\dagger}\cdot{\mathbf I}_{L\kbar\jbar}
    =2\Psi_{Rjs}^{\dagger}\Psi_{L\jbar s}\Psi_{Rk\bar{s}}^{\dagger}\Psi_{L\kbar\bar{s}}
  \nonumber\\
  -2\Psi_{Rjs}^{\dagger}\Psi_{L\jbar s}\Psi_{Rks}^{\dagger}\Psi_{L\kbar s}
    -4\Psi_{Rjs}^{\dagger}\Psi_{L\jbar\bar{s}}\Psi_{Rk\bar{s}}^{\dagger}\Psi_{L\kbar s}.
  \label{eq:AF4Bu}
\end{eqnarray}
Defining
\begin{equation}
  M_{j}^{p}=\Psi_{Rjs}^{\dagger}\tau_{ss^{\prime}}^{p}\Psi_{L\jbar s^{\prime}}+{\rm H.c.}
\end{equation}
the Hamiltonian then takes the form
\begin{equation}
  H=H_{0}-\frac{1}{2}\sum_{i,j}g_{ij}\int dx\,{\mathbf M}_{i}\cdot{\mathbf M}_{j}.
  \label{eq:HAFMx}
\end{equation}
The 2-band couplings $f_{j\jbar}^{\rho,\sigma}$ and $u_{jj\jbar\jbar}^{\rho}$
give the contributions ${\mathbf M}_{j}\cdot{\mathbf M}_{j}$ and the 2-band couplings
$c_{j\jbar}^{\rho,\sigma}$ and $u_{j\jbar\jbar j}^{\rho,\sigma}$
lead to the products ${\mathbf M}_{j}\cdot{\mathbf M}_{\jbar}$.
The 4-band couplings result in the other products ${\mathbf M}_{i}\cdot{\mathbf M}_{j}$.
For $N$ large, all couplings take the same value $g_{ij}=g>0$ ($s$-wave AFM).

As a result, all band pairs $(j,\jbar)$ are interacting with each other. In particular,
for $N$ odd, the band $r=(N+1)/2$ is interacting with all other band pairs and there
is no qualitative difference between odd and even $N$. This contrasts
the ladder-case at energies below $E_{c}$, where only interactions within the band pairs
$(j,\jbar)$ are present, respectively for odd $N$, within the band $r$. This then leads to
an odd-even effect, i.e., the band $r$ present only for odd $N$ exhibits a gapless
spinon-mode~\cite{bib:Led1}.

It is instructive to Fourier transform Eq. (\ref{eq:HAFMx})
\begin{eqnarray}
  H&=&H_{0}-\frac{1}{2}\sum_{i,j}g_{ij}
  \nonumber\\
  &&\times\sum_{k,k^{\prime},q}\left[
    \Psi_{Ris_{1}}^{\dagger}(k)\tau_{s_{1}s_{1}^{\prime}}^{p}\Psi_{L\ibar s_{1}^{\prime}}(k+q)+{\rm H.c.}\right]
  \nonumber\\
  &&\times\left[\Psi_{Rjs_{2}}^{\dagger}(k^{\prime})\tau_{s_{2}s_{2}^{\prime}}^{p}
    \Psi_{L\jbar s_{2}^{\prime}}(k^{\prime}-q)+{\rm H.c.}\right].
  \label{eq:HAFMk}
\end{eqnarray}
The Fourier transformed 2D AFM Heisenberg Hamiltonian (large-$U$ limit) takes the form
\begin{eqnarray}
  H_{J}&=&-J\sum_{{\bf k},{\bf k^{\prime}},{\bf q}}\left(e^{iq_{x}}+e^{iq_{y}}\right)
  \nonumber\\
  &&\times\Psi_{s_{1}}^{\dagger}({\bf k})\tau_{s_{1}s_{1}^{\prime}}^{p}
    \Psi_{s_{1}^{\prime}}({\bf k}+(\pi,\pi)+{\bf q}) 
  \nonumber\\
  &&\times\Psi_{s_{2}}^{\dagger}({\bf k^{\prime}})\tau_{s_{2}s_{2}^{\prime}}^{p}
    \Psi_{s_{2}^{\prime}}({\bf k^{\prime}}-(\pi,\pi)-{\bf q}),
\end{eqnarray}
where we substituted ${\bf q}$ by ${\bf q}+(\pi,\pi)$.
Therefore, the interacting part of the Hubbard Hamiltonian (\ref{eq:HAFMk}) is basically
the Heisenberg Hamiltonian restricted to the umklapp surface (the operators
$\Psi_{Rjs}^{\dagger}$ and $\Psi_{L\jbar s}$ have a momentum difference of $\pi$).
Since $U/t$ is small (and therefore the energies low), there is no $q$-dependent
coupling in Eq.~(\ref{eq:HAFMk}). In real space, this corresponds to long-range
spin-spin interactions. Note that the generalization/crossover of the Hamiltonian
(\ref{eq:HAFMk}) to 2D is straightforward. Introducing a sum over $q_{y}$
is sufficient ($q=q_{x}$).

To conclude this section: The half-filled $N$-leg Hubbard Hamiltonian --- with on-site
repulsion between electrons --- becomes at energies below the gap $\Delta$ a
Hamiltonian with purely long-range AFM spin-spin interactions.

\subsection{Physical properties: Bosonization}

Using bosonization techniques, we derive the physical properties of the
AFM Hamiltonian, Eq.~(\ref{eq:HAFMx}).

\subsubsection{Bosonized Hamiltonian}

The noninteracting part is given by $\tilde{H}_{0}$, see Eq.~(\ref{eq:H0Bos}). 
At the AFM fixed-point, the interacting part is as follows. The 4-band
interactions take the form,
\begin{equation}
  {\mathcal H}_{jk}^{4B}=
   -g_{jk}\,{\mathcal H}_{C}\left(2{\mathcal H}_{S1}+{\mathcal H}_{S2}\right)
   +(j\leftrightarrow\jbar,\,k\leftrightarrow\kbar).
\end{equation}
The ${\mathcal H}_{S1}$ contains the spin-fields $\Phi_{\sigma j\jbar-}$ and 
$\theta_{\sigma j\jbar+}$,
\begin{equation}
  {\mathcal H}_{S1}=\cos[\beta(\Phi_{\sigma j\jbar-}-\Phi_{\sigma k\kbar-}
                       +\theta_{\sigma k\kbar+}-\theta_{\sigma j\jbar+})],
\end{equation}
where $\beta=\sqrt{\pi}$ and ${\mathcal H}_{S2}$ contains the spin-fields
$\Phi_{\sigma j\jbar+}$ and $\theta_{\sigma j\jbar-}$,
\begin{eqnarray}
  {\mathcal H}_{S2}=\cos[\beta(\theta_{\sigma j\jbar-}-\theta_{\sigma k\kbar-}
              +\Phi_{\sigma k\kbar+}-\Phi_{\sigma j\jbar+})]
  \nonumber\\
  +\cos[\beta(\theta_{\sigma j\jbar-}+\theta_{\sigma k\kbar-}
              -\Phi_{\sigma k\kbar+}-\Phi_{\sigma j\jbar+})].
\end{eqnarray}
The charge-part, ${\mathcal H}_{C}$, includes the charge-fields
$\Phi_{\rho j\jbar+}$ and $\theta_{\rho j\jbar-}$,
\begin{eqnarray}
  {\mathcal H}_{C}&=&\cos[\beta(\Phi_{\rho k\kbar+}-\Phi_{\rho j\jbar+}
         +\theta_{\rho j\jbar-}-\theta_{\rho k\kbar-})]
  \nonumber\\
  &&+\cos[\beta(\Phi_{\rho k\kbar+}+\Phi_{\rho j\jbar+}
         -\theta_{\rho j\jbar-}-\theta_{\rho k\kbar-})].
  \label{eq:Cp4B}
\end{eqnarray}

In ${\mathcal H}_{C}$, the first cosine comes from the 4-band non-umklapp
interactions (\ref{eq:AF4Bnu}), while the second cosine results from the 4-band
umklapp interactions (\ref{eq:AF4Bu}). The bosonized form of the 2-band non-umklapp
terms is
\begin{eqnarray}
  {\mathcal H}_{j,1}^{2B}&=&g_{j}\cos(2\beta\theta_{\rho j\jbar-})
    \left[2\cos(2\beta\theta_{\sigma j\jbar-})\right.
  \nonumber\\
  &&\left.-2\cos(2\beta\Phi_{\sigma j\jbar+})-4\cos(2\beta\Phi_{\sigma j\jbar-})\right]
  \nonumber\\
  &&+2g_{j}\cos(2\beta\Phi_{\sigma j\jbar+})\cos(2\beta\theta_{\sigma j\jbar-})
  \label{eq:TB1}
\end{eqnarray}
and of the 2-band umklapp terms
\begin{eqnarray}
  {\mathcal H}_{j,2}^{2B}&=&g_{j}\cos(2\beta\Phi_{\rho j\jbar+})
    \left[2\cos(2\beta\theta_{\sigma j\jbar-})\right.
  \nonumber\\
  &&\left.-2\cos(2\beta\Phi_{\sigma j\jbar+})-4\cos(2\beta\Phi_{\sigma j\jbar-})\right]
  \nonumber\\
  &&-3g_{j}\cos(2\beta\Phi_{\rho j\jbar+})\cos(2\beta\theta_{\rho j\jbar-}).
  \label{eq:TB2}
\end{eqnarray}
The LLPs at the AFM fixed-point take the form
\begin{equation}
  K_{\rho j\pm}=\sqrt{\frac{\pi v_{j}\mp 3g_{j}/4}{\pi v_{j}\pm 3g_{j}/4}}
  \label{eq:AFKr}
\end{equation}
(implying $K_{\rho j+}<1$ and $K_{\rho j-}>1$) and
\begin{equation}
  K_{\sigma j\pm}=\sqrt{\frac{\pi v_{j}\pm g_{j}/4}{\pi v_{j}\mp g_{j}/4}},
  \label{eq:AFKs}
\end{equation}
i.e., $K_{\sigma j+}>1$ and $K_{\sigma j-}<1$. Table~II gives an overview about
the LLPs in the AFM and ISL phase.

Note that the commutation relation for the
field and its dual field \emph{hinders} the pinning, i.e., the localization of the field
$\Phi_{\alpha}$ and its dual field $\theta_{\alpha}$ in the minimum of a cosine at the
same time.

\begin{table}
\begin{center}
\caption{The phase at a particular energy-scale is related to the value
of the LLPs at this scale. The table shows the LLPs in the insulating
AFM, ISL, and superconducting (SC) phase. Note that always $K_{\rho j-}>1$
and $K_{\sigma j-}<1$. The differences are, that $K_{\rho j+}<1$ for the
AFM and ISL and $K_{\sigma j+}<1$ for the ISL and the SC.
}
\vspace{3mm}
\begin{tabular}{l c c c c}
      & $K_{\rho j+}$ & $K_{\rho j-}$ & $K_{\sigma j+}$ & $K_{\sigma j-}$ \\
\hline
AFM & $<1$      & $>1$      & $>1$        & $<1$  \\
ISL & $<1$      & $>1$      & $<1$        & $<1$  \\
SC  & $>1$      & $>1$      & $<1$        & $<1$  \\
\end{tabular}
\end{center}
\end{table}

\subsubsection{Charge-sector}

For the charge-sector, we then find, that the same fields are pinned as in the
half-filled two-leg (respectively $N$-leg) Hubbard ladder~\cite{bib:Led1,bib:LBFso8},
i.e., $\Phi_{\rho j\jbar+}\approx 0$ and $\theta_{\rho j\jbar-}\approx 0$. The type of
Mott insulator is thus the same in 1D and 2D. Note that pinning of
$\Phi_{\rho j\jbar-}$ instead of $\theta_{\rho j\jbar-}$ leads to another type of
insulator. The difference between these two types of insulators becomes physically
relevant upon (hole) doping. While the second type most likely becomes a FL, the
first type becomes (in case we have a spin-gap) a superconductor, since doping
only depins the $\Phi_{\rho}$-fields but does not destroy the phase coherence, i.e., the
$\theta_{\rho}$-fields remain pinned, see~Ref.~\cite{bib:Led1}.

\subsubsection{Spin-sector}

In contrast, in the spin-sector, both the fields $\Phi_{\sigma j\jbar\pm}$ and the dual
fields $\theta_{\sigma j\jbar\pm}$ appear in a cosine, resulting in a competition between
different ``phases'', i.e., pinning of the field vs. its dual field. However, since
$K_{\sigma j+}>1$ and $K_{\sigma j-}<1$ (for a comparison, see Sec.~III~A), it is more
favorable to pin $\theta_{\sigma j\jbar+}$ and $\Phi_{\sigma j\jbar-}$ than the
corresponding dual fields.

The physical interpretation is then the following. The pinning of $\Phi_{\sigma j\jbar-}$
leads to a spinon confinement, leaving as physical particles spin~1 magnons. Note that
only the \emph{differences} $\theta_{\sigma j\jbar+}-\theta_{\sigma k\kbar+}$ appear in
$H^{4B}$, such that the total magnon mode(s), given by $\theta_{T}=\sqrt{2/N}\sum_{j}\theta_{\sigma j\jbar+}$
and $\Phi_{T}=\sqrt{2/N}\sum_{j}\Phi_{\sigma j\jbar+}$, remains gapless. The $\pi$ and $0$
mode then result from a superposition of left/right going modes along the chains (since
we use open boundary conditions, the transvers momentum is always positive).

For the calculation of the spin-spin correlation function, we first rewrite the
real-space operators $d_{is}$ in terms of the band operators $\Psi_{hjs}$ and
then the band operators in terms of the bose-operators. Using that only products
of fields, which contain the pinned charge-fields $\Phi_{\rho j\jbar+}$ and
$\theta_{\rho j\jbar-}$ give non-vanishing contributions to the spin-spin
correlation function, the real-space spin-operator at the position $(x,i)$ takes
the form
\begin{eqnarray}
  S_{i}^{p}(x)=\frac{1}{2}\sum_{j}\gamma_{ij}\gamma_{i\jbar}
    \bigg(\Psi_{Rjs}^{\dagger}\tau_{ss^{\prime}}^{p}\Psi_{L\jbar s^{\prime}}
    e^{-i(k_{Fj}+k_{F\jbar})x}
  \nonumber\\
    +R\leftrightarrow L\bigg),
  \label{eq:SCf}
\end{eqnarray}
where $\gamma_{jm}=\sqrt{2/(N+1)}\sin[\pi jm/(N+1)]$ and at half-filling,
$k_{Fj}+k_{F\jbar}=\pi$. The abelian bosonization scheme used here, \emph{breaks}
non-abelian symmetries, i.e., the SU(2) spin-symmetry is broken down to U(1), see,
e.g., Refs.~\cite{bib:Voit,bib:Wit}. In our case, only the $x$ and $y$ components
of the spin-spin correlation function give then straightforwardly the correct
physical result (this ``problem'' occurs also in a single chain with a
spin-gap~\cite{bib:Voit}). The products of fields appearing in Eq.~(\ref{eq:SCf})
are then rewritten in terms of bose-operators according to
\begin{equation}
  \Psi_{Rj\uparrow}^{\dagger}\Psi_{L\jbar\downarrow}
    \propto e^{i\sqrt{\pi}(-\Phi_{\rho j\jbar+}-\Phi_{\sigma j\jbar-}
    +\theta_{\rho j\jbar-}+\theta_{\sigma j\jbar+})}.
\end{equation}
The charge-fields and $\Phi_{\sigma j\jbar-}$ are pinned and can be set to zero. Using that
$\theta_{\sigma j\jbar+}-\theta_{\sigma k\kbar+}\approx 0$, we express $\theta_{\sigma j\jbar+}$
in terms of the total spin-mode $\theta_{T}$, $\theta_{\sigma j\jbar+}=\sqrt{2/N}\theta_{T}$.
Since $H_{0}$ is gaussian in the fields $\Phi_{T}$ and $\theta_{T}$, 
the spin-spin correlation function takes the form (for details about such calculations, consult,
e.g., Ref.~\cite{bib:Gogolin})
\begin{equation}
  \langle{\bf S}_{i}(x)\cdot{\bf S}_{l}(0)\rangle\propto(-1)^{i+l}\cos(\pi x)/x^{1/N},
\end{equation}
where $x<\xi_{c}\propto 1/E_{c}$ (for $x>\xi_{c}$ we are in the ISL phase, where the decay
is exponentially with coherence length $\xi_{AF}\sim\xi_{c}$, respectively, for $N$ odd,
$\propto 1/x$).

The energy-scales for charge and spin-excitations are \emph{different}. The
LLPs in the charge-sector, see Eq.~(\ref{eq:AFKr}), deviate more from their
noninteracting value ($=1$) than the LLPs in the spin-sector, Eq.~(\ref{eq:AFKs}), i.e.,
\begin{equation}
  |K_{\rho j\pm}-1|\approx 3g_{j}/(4\pi v_{j}),\;\;|K_{\sigma j\pm}-1|\approx g_{j}/(4\pi v_{j}).
\end{equation}
This implies that the charge-fields are pinned more strongly than the spin-fields and therefore
the Mott-gap (we may interpret the RG-scale $\Delta$ as the charge-gap) is larger than the gap
to spin-1/2 (spinon) excitations. The coupling between the different bands
makes it difficult to investigate the detailed excitation spectrum, such that we
do not go beyond this qualitative result (for the pure sine-Gordon model, there exist
rigorous results about the gap dependence on $K_{\alpha}$, see Ref.~\cite{bib:Gogolin}).

\section{Effect of (hole) doping}

Next, we study the effect of doping away from half-filling. There are two \emph{different}
doping regimes. The \emph{lightly doped} case, where the influence of the umklapp
processes has to be taken into account, can be treated as a perturbation of the
half-filled low-energy Hamiltonian (limit $\delta\rightarrow 0$ and $U$ finite). For
\emph{increasing} doping, one has then again to investigate the RG flow.

For the lightly doped case (Sec.~V~A), we find above $E_{c}$ the same phase as below
$E_{c}$ (see Ref.~\cite{bib:Led1}): a conducting spin-liquid (bound hole-pairs), see
also Fig.~\ref{f:sphd}. Furthermore, while at half-filling the magnetic correlations
are only at $(\pi,\pi)$, they become \emph{in}commensurate upon doping.

In Sec.~V~B, we briefly show how and why for increasing doping phase-coherence between all
band pairs sets in and the system becomes a 2D-like $d$-wave superconductor. The low-energy
Hamiltonian takes the form of a $d$-wave BCS model, where the pairing is between electrons.

Note that the superconducting instability is for $N\gg 2$ generated in the same
way as for the 2D case~\cite{bib:ZS,bib:HaMe,bib:HSFR}: A Kohn-Luttinger-type
attraction~\cite{bib:KoLu} is mediated by AFM processes~\cite{bib:SLH}. In contrast, the
superconducting instability in the two-leg ladder is generated by forward-scattering
processes~\cite{bib:LBF}.

\subsection{Lightly doped case}

The ISL phase of the half-filled two-leg Hubbard ladder becomes superconducting
when doped, because of the phase-coherence (see discussion of the charge-sector
in the previous section) and the spin-gap present at half-filling~\cite{bib:LBFso8}.
Here, we show that the doped Hubbard AFM becomes a conducting spin-liquid, formed
by bound hole-pairs (this phase may becomes superconducting at very low energies;
for the 4-leg ladder, see discussion in Ref.~\cite{bib:Led1}).

The \emph{2-band} interactions alone, see Eqs.~(\ref{eq:TB1}) and (\ref{eq:TB2}),
would result in an ISL phase, i.e., $\Phi_{\sigma j\jbar\pm}\approx 0$. It is
the presence of interactions between bands $i$ and $j$ which are \emph{not}
``paired'' ($j\neq i,\ibar$), which renders the system a 2D-like AFM. The effect
of a decreasing $N<\infty$ is therefore that interactions between unpaired
bands become suppressed and the AFM correlations are reduced. Next, we show that the
effect of doping away from half-filling is a similar one, i.e., the doping
$\delta$ corresponds to $1/N$.

At low dopings and at finite $N$, we can dope the system perturbatively by introducing
a chemical potential term $-\mu Q$ ($Q$ is the total charge) in the half-filled
low-energy Hamiltonian, see Ref.~\cite{bib:Led1}. For the charge-sector, we can use
bosonization, where the charge density is given by
$\rho=\sum_{j}\partial_{x}\Phi_{\rho j}$.

\begin{figure}[t]
  \centerline{
   \psfig{file=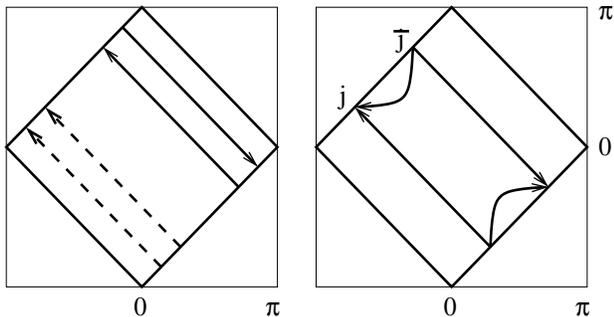,width=8.1cm,height=4.2cm}}
  \vspace{1mm}
  \caption{
  The square is the umklapp surface, which is the FS at half-filling ($t=t_{\perp}$).
  Left: There are 2 types of AFM processes; AFM processes
  which are umklapp processes (dashed arrows), and AFM processes which are
  not umklapp processes (solid arrows). Right: The (non-umklapp) AFM
  processes which take place within a band pair $(j,\jbar)$ are identical to
  the corresponding ($d$-wave) Cooper processes within this band pair.
  Upon doping away from half-filling, these processes remain
  large, open a spin-gap and lead to phase coherence within band pairs
  $(j,\jbar)$. Note that we use for the ladders open boundary conditions,
  whereas the above figures are drawn --- for a better comparison with the 2D
  case --- for periodic boundary conditions.
  } 
  \label{f:afp}
\end{figure}

Doping then introduces \emph{kinks} in the fields $\Phi_{\rho j\jbar+}$, such that
the expectation value of cosine terms, which contain this field gradually goes
to zero (the kink is a fermionic, strongly localized particle). Here, the kinks are
bound hole-pairs of charge~2 and zero spin. All 4-band interactions
contain the field $\Phi_{\rho j\jbar+}$, see Eq.~(\ref{eq:Cp4B}), and therefore
vanish. The effect of the doping on the 2-band interactions is the same as when
doping an ISL, i.e., in the bosonized version, the umklapp term (\ref{eq:TB2})
vanishes leaving only the term (\ref{eq:TB1}), which opens a spin-gap, and leads to
$d$-wave-like phase coherence between the bands $j$ and $\jbar$, corresponding to
the Fermi momenta $(k_{Fj},\pi-k_{Fj})$ and $(\pi-k_{Fj},k_{Fj})$. The charge-gaps
close to $(\pi/2,\pi/2)$ are the smallest ones (note that
$g_{N/2}/v_{N/2}<\ldots<g_{1}/v_{1}$), such that the hole-pairs enter first there.

In other words, it is the fact, that the ($s$-wave) AFM processes
\begin{equation}
  -{\bf M}_{j}\cdot{\bf M}_{\jbar}=
    -\Psi_{Rj\downarrow}^{\dagger}\Psi_{L\jbar\uparrow}\Psi_{Lj\uparrow}^{\dagger}\Psi_{R\jbar\downarrow}
    \;(+{\rm other\;terms})
\end{equation}
(partially) coincide with the ($d$-wave) Cooper processes with momentum transfer $\pi$
\begin{equation}
  \Delta_{j}^{\dagger}\Delta_{\jbar}=
    \Psi_{Rj\downarrow}^{\dagger}\Psi_{Lj\uparrow}^{\dagger}\Psi_{L\jbar\uparrow}\Psi_{R\jbar\downarrow}
    \;(+{\rm other\;terms}),
\end{equation}
which makes a doped AFM a superconductor: These particular AFM/Cooper processes flow at higher
energies --- driven by AFM processes --- to strong coupling and are not suppressed at lower
energies by doping (see also Fig.~\ref{f:afp}). We therefore have AFM mediated
superconductivity~\cite{bib:SLH}. The full phase-coherence around the FS then grows out of the
phase-coherence between bands $j$ and $\jbar$, see below.

Upon doping, the AFM correlations become \emph{in}com\-mensurate. When doping away from
half-filling, $k_{Fj}+k_{F\jbar}<\pi$, and the AFM peak shifts away from $(\pi,\pi)$,
see Eq.~(\ref{eq:SCf}). In particular, for small $t_{\perp}/t\ll 1$, all bands are
(almost) equally doped and $k_{Fj}+k_{F\jbar}=\pi(1-\delta)$, i.e., the incommensurability
is equal the doping $\delta$. Furthermore, doping leads to a
cutoff of the AFM processes at the energy-scale of the chemical potential $\mu$
and therefore to an exponential decay of the spin-spin correlation function,
where the coherence-length is given by $\xi_{AF}\propto 1/\mu$.

\subsection{Increasing doping}

We argue, that for increasing doping, phase-coherence between the band pairs sets in,
when the ``distance'' between neighboring bands is of the order of the ``distance'' of
the FS to the umklapp surface, i.e., for dopings $\delta>\delta_{c}(N)\sim(t_{\perp}/t)/N$.
The system then becomes a 2D-like $d$-wave superconductor.

Since a finite chemical potential $\mu$ results in a low-energy cutoff for the umklapp
and 4-band-AFM interactions, one has to start initially with all interactions which
are \emph{large} at this cutoff energy-scale. In leading order, this implies that not
only the 4-band interactions $c_{jk\kbar\jbar}^{\rho,\sigma}$ and
$u_{jk\kbar\jbar}^{\rho,\sigma}$ have to be taken into account, but also ``neighbor''
processes of the form $c_{jk\kbar\pm 1\jbar\pm 1}^{\rho,\sigma}$ and
$u_{jk\kbar\pm 1\jbar\pm 1}^{\rho,\sigma}$. These processes are only present down to an
energy-scale $t_{\perp}/N$. For dopings $0<\delta<(t_{\perp}/t)/N$, the ``neighbor''
processes are cutoff \emph{before} the main AFM processes, which leads to a decoupling
into band pairs $(j,\jbar)$, while for $\delta>(t_{\perp}/t)/N$ all AFM processes are
cutoff by the chemical potential. In this case, not only 2-band Cooper interactions
within band pairs $(j,\jbar)$ are large at the cutoff energy-scale, but also the
neighboring 2-band Cooper interactions $(j,\jbar\pm 1)$.

Integrating the RGEs without AFM and umklapp processes, but with initial-values for the
Cooper and forward interactions given by the AFM phase, see Eq.~(\ref{eq:Ratios2B}), and
similar initial-values for non-leading Cooper processes $c_{j\jbar\pm p}^{\rho,\sigma}$
($p=1,2,3,\ldots$), we obtain a $d_{x^{2}-y^{2}}$ superconducting instability at lower
energies. We always find that the scaling of the couplings is exactly such that (in
particular, $4c_{ij}^{\rho}=c_{ij}^{\sigma}$, $f_{ij}^{\rho}\approx f_{ij}^{\sigma}\approx 0$)
\begin{equation}
  H=H_{0}+\sum_{i,j}\int dx\,V_{ij}\Delta_{i}^{\dagger}\Delta_{j},
\end{equation}
where the pairing operator is
$\Delta_{j}=\Psi_{Rj\uparrow}\Psi_{Lj\downarrow}+\Psi_{Lj\uparrow}\Psi_{Rj\downarrow}$
and $V_{ij}<0$ for $i,j\leq N/2$ respectively $i,j>N/2$ and $V_{ij}>0$ in all other cases
($d_{x^{2}-y^{2}}$ symmetry). The size (but not the sign) of $V_{ij}$ depends on the
initial-values (i.e., on the doping). Note that the energy-scale of the system is reduced
by doping, i.e., the superconductor has a lower scale than the AFM.

For large dopings, the 4-band interactions are cutoff \emph{before} they
could have flown sufficiently close to the AFM fixed-point, such that there is no
instability in the RG flow and the low-energy phase is a FL~\cite{bib:LBF}.

As a result, for $N$ finite, there is between the insulating AFM respectively ISL
phase at half-filling and the $d$-wave superconductor at higher doping, an
intermediate, conducting phase with a spin-gap (see also Fig.~\ref{f:sphd}).
In the 2D-limit, the intermediate phase seems to disappear.

\section{Conclusions}

We have studied the dimensional crossover from 1D to 2D in half-filled, weakly
interacting, $N$-leg Hubbard ladders for $t_{\perp}<t$. Using the RG method,
we found a crossover energy $E_{c}\sim t\exp[-\alpha\exp(\gamma N)]$, where at
energies below $E_{c}$ the behavior of the spin-sector is ladder-like and at
energies above $E_{c}$ 2D-AFM-like. We showed that the 4-band-AFM processes are
responsible for an instability in the RG flow. Since without 4-band and umklapp
processes the Hubbard ladders become a FL for large $N$~\cite{bib:LBF}, AFM
processes (at higher energies) are necessary and sufficient for a nontrivial
phase in the (doped) $N$-leg Hubbard model.

In the AFM phase, the couplings flow towards universal ratios and we obtained an
analytical expression for the Hamiltonian which is similar to the Heisenberg
AFM. Using bosonization techniques, we further investigated this Hamiltonian and
found that the charge-sector is the same as in the half-filled two-leg ladder,
i.e., the type of Mott insulator is the same in 1D and 2D. For the spin-sector,
bosonization techniques confirmed that the small-$U$ case has similar physical
properties as the large-$U$ (Heisenberg) case.

The effects of doping on the half-filled Hubbard Hamiltonian and of reducing the
number of coupled chains $N$ is (almost) the same: In both cases, the interactions
between unpaired bands are suppressed, the antiferromagnetic correlations are
reduced and only interactions within band pairs $(j,\jbar)$ remain. Doping then
opens a spin-gap, but phase coherence exists only between bands $j$ and $\jbar$.
At higher doping levels $\delta\sim(t_{\perp}/t)/N$, we can expect that phase
coherence between all bands sets in, such that the system becomes a 2D-like
$d$-wave superconductor. While the lightly doped case is conveniently described
by bound hole-pairs, the 2D-like superconductor at higher dopings is represented
by a BCS-like $d$-wave Hamiltonian with electron pairing. In the doped case the
antiferromagnetic correlations become incommensurate.

As a result, the Hubbard model cannot be understood in terms of an AFM, ISL or
superconductor alone. The phase (correlation function) depends always on the
energy-scale (respectively length-scale) and is related to the value of the
LLPs. The AFM exhibits the highest energy-scale. At half-filling, the
groundstate (for $N$ finite) is the ISL and for sufficient doping the $d$-wave
superconductor. We conjecture that in the doped case, the flow of the couplings
from the AFM to the $d$-wave takes place via an \emph{intermediate} phase:
Consult table~II; AFM and superconductor differ in \emph{two} LLPs. This
intermediate (finite energy) phase has then the same LLPs as the ISL ---
suggesting that the spin-gap opens \emph{before} phase coherence sets in.

We believe, that also for other systems of interacting electrons the
dimensional crossover from 1D to 2D can be studied in a similar way and maybe
enlightens our understanding of the 2D case.

\acknowledgments{I am grateful to B. Binz, C. Honerkamp, K. Le Hur, T.M. Rice, and
M. Sigrist for fruitful discussions and comments throughout this work.}

\end{multicols}

\end{document}